# 3D Modelling to Address Pandemic Challenges: A Project-Based Learning Methodology

Tânia Rocha[1,2][0000-0002-2605-9284], Ana Ribeiro[1], Joana Oliveira[1], Ricardo Nunes[1,2][0000-0002-7557-2121], Diana Carvalho[1,2][0000-0001-9736-7893], Hugo Paredes[1,2][0000-0002-4274-4783] and Paulo Martins[1,2][0000-0002-3040-9080]

[1] UTAD - University of Trás-os-Montes e Alto Douro, Quinta de Prados, 5001-801 Vila Real, Portugal
[2] INESC TEC, Rua Dr. Roberto Frias, 4200-465 Porto, Portugal
trocha@utad.pt, {al77276, al75499}@alunos.utad.pt, {rrnunes, dianac,hparedes,pmartins}@utad.pt

**Abstract.** The use of 3D modelling in medical education is a revolutionary tool during the learning process. In fact, this type of technology enables a more interactive teaching approach, making information retention more effective and enhancing students' understanding. 3D modelling allows for the creation of precise representations of the human body, as well as interaction with three-dimensional models, giving students a better spatial understanding of the different organs and systems and enabling simulations of surgical and technical procedures. This way, medical education is enriched with a more realistic and safe educational experience. The goal is to understand whether, when students and schools are challenged, they play an important role in addressing health issues in their community. School-led projects are directed towards educational scenarios that emphasize STEM education, tackling relevant public health problems through open-school initiatives. By implementing an educational scenario focused on 3D modelling and leveraging technology, we aim to raise community awareness on public health issues.

## 1    Introduction

As technology continues to advance, virtual and augmented reality, as well as 3D modelling, are becoming increasingly relevant across various sectors, especially healthcare. The use of 3D models has been gaining importance, being applied both to improve surgical training and to create more effective treatment plans. Furthermore, it plays a crucial role in the development of assistive technologies such as prosthetics and orthoses, as well as tools and materials that support specific medical activities.

3D technology is revolutionizing the way products are designed, manufactured, and managed, offering numerous advantages. These include significant improvements in the quality and efficiency of patient care, while also serving as a teaching resource for professionals at all stages of their careers, from students to interdisciplinary teams. It also aids in planning medical interventions, identifying problems, demonstrating them to healthcare professionals, and providing more effective patient follow-up. Given its impact, we believe it is essential to include the potential of 3D modelling as a key topic in the science and technology curriculum.

The COVID-19 pandemic highlighted global health challenges, further exacerbated by the population's low health literacy, underscoring the need for enhanced public health education. This education has the power to raise awareness about risk factors, health behaviors, and the social and environmental determinants of health. In this context, we propose developing and implementing an educational scenario focused on 3D modelling, using a project-based learning approach to explore these issues. This would strengthen the capacity of students and schools to promote learning in Science, Technology, Engineering, and Mathematics (STEM), with a particular emphasis on public health issues.

The educational scenario aims at supporting 8th-grade Science and ICT teachers in using 3D environments, drawing on up-to-date scientific and technical evidence. This learning experience will enable students to develop a deeper understanding on how STEM fields can be used to address public health challenges, encouraging informed personal decision-making and promoting public policy development. Critical thinking, essential for effective decision-making, is fostered when students can engage in real-world problem-solving processes, and project-based learning offers a valuable pathway to achieve this.

Ultimately, our study seeks to involve students and schools in preparing their communities for major health challenges. By implementing an educational scenario focused on 3D modelling, we intend to maximize the use of this technology and raise awareness about public health issues, encouraging greater participation and action by the community.

## 2    Background Overview

3D modelling has had a very positive impact on STEAM education and in public health [1, 2, 3]. In the teaching area, it has been a challenge to train 3D modelling skills [1] to obtain acceptable results. However, we have often found that the use of 3D modelling encourages students to expand their awareness, making knowledge more accessible to teachers [4]. STEAM education with modelling tends to be innovative and allows students to accelerate their acquaintance with the field in a simple way [5]. In the health area, as with STEAM teaching, 3D modelling proves to be a useful tool as it allows for deeper teaching. [3] This technology allows us to visualize anatomical models, simulate procedures and visualize viruses or other threats to human health, facilitating visualization as well as communication between professionals in the field [6].

As such, 3D modelling emerges as an indispensable tool in both areas of study, as it enables future generations to face future challenges [7].

## 2.1 Project-Based Learning

Project-based learning (PBL) is an educational method where students engage in individual and group projects to explore real-world challenges. This approach helps students understand the relevance of the content and how it can be applied in practical, meaningful ways [8].

Extensive evidence shows that PBL is an effective approach for teaching students' complex skills and processes, including planning, communication, problem-solving, and decision-making [3]. Critical thinking development is enhanced when students engage in problem-solving and knowledge creation in real-world situations. Evidence shows that PBL is an effective approach to support this process [9].

This learning model offers a broad platform for developing 21st-century skills, such as critical thinking, problem-solving, communication, collaboration, creativity, networking, and the effective use of technology [10,11].

It incorporates real-life challenges, focusing on authentic (not simulated) problems or questions, where the proposed solutions have the potential to be applied in real-world situations [3]. These projects involve complex tasks centered on challenging questions or problems that engage students in activities like design, problem-solving, decision-making, or investigation. They provide students with the opportunity to work independently over extended periods and result in the creation of realistic products or presentations [3].

Indeed, it is significantly more effective than traditional instruction for training competent and skilled practitioners, while also promoting long-term retention of the knowledge and skills gained during the learning experience or training session [12]. This learning method involves an instructional approach where students tackle real-world challenges through inquiry-based learning, leading to the completion of meaningful projects and active knowledge construction. Inquiry-based skills are essential for individuals to think independently and critically engage with publicly available information on important topics [9].

Project-based learning emphasizes student autonomy and collaborative learning [13], which have been found to improve motivation and attitude [14]. Students can exercise autonomy throughout the process, making their own decisions. This autonomy enhances their motivation, strategic thinking, and ability to anticipate outcomes [13]. Implementing project-based learning in educational settings presents challenges: students are required to take ownership of their learning process while receiving adequate support and tools to develop their projects. Therefore, the learning environment and teaching practices must be intentionally designed to foster and support students' self-regulated learning [9].

Furthermore, parents, educators, and policymakers are increasingly recognizing that effective project-based learning enhances student engagement—an aspect that has gained particular importance in light of the disruptions caused by the pandemic—and helps connect academic content to the wider world beyond the classroom [8].

Project-based learning (PBL) encompasses a driving question, fosters collaboration among students, utilizes scaffolding technologies, emphasizes critical thinking and communication skills, and promotes interdisciplinary learning [9]. The shift in expertise from teacher-directed, assigned work focused on comprehension and assimilation to a student-directed, goal-oriented learning model can significantly enhance the current state of research on "knowledge transfer" [14].

In the proposed educational scenarios, students engage in project work that extends beyond the information provided in the classroom, allowing learning to occur in a broader context. They share the outcomes of their projects with peers, parents, local community members, and community leaders, as well as policymakers [9].

## 3   Educational Scenario

The educational scenario serves as a comprehensive instructional and learning framework designed to effectively structure the teaching-learning process. By integrating inquiry-based and project-based learning as fundamental approaches for skill development, it actively involves students, teachers, and the wider school community in research projects focused on a vital public health issue: utilizing 3D modelling to address challenges related to pandemics.

This scenario specifies clear learning objectives, provides digital learning resources, and includes a detailed teaching script that features lesson plans, assessment methods, supplementary activities, and guidelines for school-based projects on the topic. Additionally, the script facilitates connections between STEM professionals—such as researchers, public health specialists, engineers, and project managers—and the classroom, enabling students to take charge of their projects and present their results during an open-school event.

### 3.1   Target Audience

The educational scenario was developed to $8^{th}$ graders students (14-year-old students). Science and ICT classes are preferred to be the ground for the scenario enactment. However, it can be implemented not only by ICT teachers, but also these teachers can integrate other colleagues in the enactment of the scenario (e.g., ICT, visual education, mathematics and English teachers), as it aims to be interdisciplinary.

### 3.2   Learning Goals/Competences

There are specific key STEM-related competences worth of mentioning, as well as those connected with 3D modelling and innovation. The learning goals and expected overall outcome assessment were if students are able to:
1. Use online tools for 3D modelling.
2. Analyze pre-designed models.
3. Identifies 3D environments and basic features.
4. Designs basic shapes and elements in a 3D environment.
5. Export modelling objects.
6. Describes different approaches to create 3D objects for positively influencing global health.
7. Gives examples of how 3D models can contribute to improve healthcare environments.

Overall, the educational scenario expects key competences acquired within three scopes: knowledge, skills (abilities/competences), beliefs (attitudes/behavior). Specifically:

**Knowledge.** Expected outcome assessment regarding overall 3D modelling concepts:
1. Understands the 3D technical principles and workflows.
2. Recognizes software basic features regarding the interface.
3. Recognizes software basic features regarding shapes.
4. Recognizes software basic features regarding textures and illumination.
5. Recognizes software basic features regarding rendering.
6. Is able to understand the importance of 3D environments to address pandemic challenges and ensure public health.
7. Is able to understand the importance of 3D environments in the health care industry in order to decrease inequality and improve inclusion.

**Skills.** Expected outcome assessment regarding 3D modelling basics, imagination, creativity:
1. Designing 3D elements by combining process knowledge, computational design tools, and application requirements.
2. Technical usage of 3D software.
3. Recognizes appropriate proficiencies necessary for 3D modelling.
4. Is able to understand the virtual environment.
5. Can create specific 3D objects and sets.
6. Is able to identify the differences of multiple 3D modelling software.

**Beliefs.** Expected outcome assessment regarding affective, attitudes and behaviors:
1. Using imagination for designing real tools and materials.
2. Using creativity skills on new technologies in the development process of the solution.

**Attitudes and behavior.** Specific outcome assessment.
1. Recognizes the importance of raising awareness on how 3D modelling can help the community.
2. Has intention to continue extending the skills and knowledge regarding 3D modelling?
3. Is aware of the democratization of 3D modelling?
4. Has a positive attitude towards 3D modelling?
5. Believes that is important to improve one's own personal capabilities regarding 3D modelling.

### 3.3 Estimated duration

The scenario is divided in different sessions and scope: 7 inside-classroom sessions of 40-45 min each (lesson 1 - lesson 7), plus 4 outside-classroom sessions of 40-45 min each (lesson 8 - lesson 11), for supplementary learning activities and the school project.

### 3.4 Teaching-Learning Activities (Lesson Plan)

Next, each lesson plan of the scenario is presented.

**Lesson 1.** Introduction of virtual environments.
The teaching-learning script started with a question "what is a virtual environment (V.E.)"?

For that, a brainstorming session on the questions: "what is a virtual environment?" and "how can modelling be a convergence point for STEM?" is done.

Then, students were divided into groups and asked to Google key definitions of virtual environments and their impact on STEM. Each group produced at least three different sentences; read them and selected the main keywords for sharing. Then, students were asked to go to the flipchart or whiteboard and write the main keywords selected.

The next step was a video presentation about virtual environments. After, a discussion was mandatory about their previous definitions and keywords and their recent new knowledge about the topic learned.

**Lesson 2.** The benefits of 3D modelling in healthcare during / after a pandemic event.
In lesson 2, after a short conversation about the previous lesson, the benefits of 3D modelling in healthcare were presented. For that, were developed specific digital educational resources presenting the benefits of 3D modelling in healthcare environments; and, an introduction of 3D modelling for product design in healthcare.

After the brainstorm on what is a virtual environment, students were provided with infographics on how these environments can contribute positively to the healthcare industry. Examples: in rehabilitation, surgical training, treatment plans, assistive technologies, (prosthetics, orthosis), product design and production, patient care.

Then, a group discussion was stimulated around the question "What did Covid-19 change in my life?" Students are asked to share their own experiences during and after the first outbreak of Covid-19. The main goal is to understand their awareness of the depth the pandemic event had in their lives and channel their responses towards the demands of the healthcare sector, to help them understand how virtual environments could help mitigate issues / challenges in healthcare.

Furthermore, a debate around the question "How can 3D modelling help with pandemic challenges?" Students were asked to break into groups and each group must provided an example on how 3D modelling can tackle one specific pandemic issue, namely identify specific products that can be modelled and produced for that end, supporting arguments and counter-arguments. Example: products for improving health care and quality of life after a pandemic event, e.g., help in the treatment of depressive symptoms, prolonged stress, anxiety, insomnia, denial, fear, and anger.

**Lesson 3.** Introduction 3D modelling and principles.

After a short conversation about the previous lesson, 3D principles and approaches were presented to be discussed. The introduction on 3D modelling with a PowerPoint presentation with several examples. Students experimented a virtual environment using a headset apparatus and proper software. Furthermore, several videos regarding 3D models and environments; and, pedagogical glossary for technical terms and definitions were provided.

Additionally, a video was presented to introduce the 6 Key principles for 3D modelling (form, detail, scale, adaptation, reuse, surface quality).

Basic variables (X, Y, Z) were presented and correlated with horizontality, verticality and depth. Simple exercises were done, and replicated by the students, demonstrating these variables.

Also, a group discussion was suggested on: "How can we design this object in 3D? E.g., surgical mask." The aim was to show different basic objects and discuss and reveal which basic elements can be used to model the objects shown. Students compared different models of the same object and be aware of: the differences they had in the meshes; what benefits and limitations each one had; what situations each model were more suitable for. Also, they recognized the limitations of scientific models and their differences between real-world objects.

**Lesson 4.** 3D modelling software basic features (Knowing the Interface).
The teaching-learning script started with the presentation of the software interface, providing an individual hands-on approach.

The digital educational resource provided was a 3D modelling tutorial about software interface (video). A video on the software's interface and major features was shown. After, individually, students replicated some basic functionalities in the computer: first approach of the software environment and features.

After this first approach, a simple tutorial was provided and students will autonomously and individually follow it, step by step.

The debate was around the questions: "What were the software presented?"; "Are there only paid software for 3D modelling?"; "Which are the major features of the software?"

**Lesson 5.** 3D modelling software basic features: SHAPES.
Students were introduced to geometric representation of models in 3D environment.

The digital educational resources provided were for learning types of shapes, an infographic, a tutorial (step by step). Also, it was encouraged group work (accord with the availability of laptops or tablets for group work).

Students were organized in groups (1 group – 1 Object) and invited to explore shapes in the creation of simple daily objects. After, they presented their work to their colleagues.

**Lesson 6.** 3D modelling software basic features (textures and illumination).
The learning objects offered were a 3D modelling tutorial about textures (video and tutorial).

Students had an overview about the application of simple textures in objects by watching a video. Then, following a step-by-step tutorial, they experimented to apply texture in objects previously modelled.

As a digital educational resource, a video on 3D modelling illumination was also run. As illumination plays a major role in realism on 3D environments, some basic aspects about illumination were presented to the students.

**Lesson 7.** 3D modelling software basic features (rendering).
To finalize the first complete exercise in 3D modelling environment, students learnt the process of RENDER. For that:
1. a manual was provided as a digital educational resource on 3D Render;
2. a questionnaire was used - as quantitative assessment – to assess impact in terms of students' knowledge, skills, attitudes and behavior;
3. and, a presentation and group activity, also as qualitative assessment.

Students presented their modelling objects in groups and, for each presentation, the other colleagues identified which features, shapes and textures were used or which other solutions may be used to improve to object presented.

**Lesson 7-forward.** The school project.
After building and presenting their work, students were challenged to model other 3D objects in groupwork. This was defined as the School Project described next.

### 3.5 School Project and Supplementary Educational Activities

The final school research project focuses on the following topics: importance of 3D modelling; technical features and principles of 3D modelling; possible applications of 3D modelling in public health. The students' main challenge defined in the phase was to model a 3D object to address communicable diseases challenges.

Supplementary educational activities are expected during this stage. devoted to the preparation of the school project, these activities include (but are not restricted to):
1. Teleconference with STEM professionals (e.g., engineers, designers, medical doctors, researchers). Students asked questions to experts with a particular focus on: a) future academic choices and career paths; b) identifying new professions in new fields of industry 4.0.
2. Visits to field-related laboratories (e.g., FABLABs). Students visited the working environment and dynamic of a FABLAB. Here, students were provided with the opportunity to make questions to experts within the laboratory, see the processes and go along with the working dynamic. These activities were relevant for students' connections with STEM curricula, careers and professionals.

This specific learning phase was dedicated to the school research project. Students were organized in groups; each group addressed one object based on the daily pandemic challenges lived. The project challenges each group of students to: 1) identify and represent their progress in the form of essay responses and using Likert scales to show their improvement from the first lesson to the last; 2) model and present an object with what they have learned throughout the teaching-learning sequences and the ideas that emerged during the teleconference with experts.

A competition and reward for the best 3D objects took place in a final open-schooling event.

Furthermore, the teaching-learning process milestones were defined:

1. Students should be able to propose solutions for 3D modelling basic objects (masks, ventilators…).
2. Students should be able to communicate the findings, motivations and limitations of various 3D elements and shapes considered in the working process.
3. Students should be able to identify and communicate the importance of 3D modelling to address pandemic challenges but also the role of Innovation.
4. Students should be able to use technical argumentation to justify policy choices.

### 3.6 Organization of the Open Schooling Event

At the end of the lessons and development of the school project, an open-schooling event is organized. Each project output (3D object) is presented by the students in a community setting (e.g., exposition center, municipality, garden, museum, science fair) in a 3D prepared environment (all apparatus included). Students will prepare a pitch on how 3D modelling can address pandemic challenges and technical speeches to motivate peers to new technologies and environments will occur. Students, parents, school community and relevant local stakeholders may attend the event and are introduced on the topic on how 3D modelling can be used to address pandemic challenges. Furthermore, the scenario has a multidisciplinary approach, such as in art, design, engineering and mathematics. The finals results of the competition and reward for the best 3D objects will take place, as well as the dissemination of evidence recommendations via social, community and conventional media. The goal is to formulate a final public debate around the topic at hand.

### 3.7 Assessment Methods

The outcome assessment is twofold: qualitative and quantitative. The qualitative assessment focuses on the students' final school project, which involves creating a 3D model within a STEM context, focusing on public health. For the quantitative assessment, a series of questionnaires will be administered to evaluate the impact on students' knowledge, skills, attitudes, and behaviors.

The evaluation of the teaching-learning sequence includes an observation grid to track several key aspects: reaching the intended audience and its scope, whether the scenario was implemented as planned, how well the learning activities unfolded, any organizational issues that need addressing, the planned duration of the sequence, and the total number of participants involved. Additionally, students' feedback will be gathered through a likeability score, asking how enjoyable the experience was, how likely they would be to repeat it, and what improvements could be made. ("how fun was it to do / how fun would be to do again / how could it be better").

## 4      Conclusions and Future Work

Considering the educational scenario of 3D modelling, both teachers and students will need to adapt to this new teaching approach, as it is not included in the national school curriculum. This approach poses a significant challenge, as it requires students to step out of their comfort zones, particularly regarding autonomy, teamwork, decision-making, and the use of technology. For teachers, this method is also demanding, as they will need to support students throughout the entire process, helping them overcome difficulties and keeping them motivated. Additionally, effective time management during class sessions and project phases will be essential.

After the theoretical lessons, students will engage in practical sessions outside the classroom, where they will develop their projects based on information gathered through teleconferences with STEM professionals and visits to laboratories. These experiences will provide a comprehensive overview of the proposed educational scenario.

In the final phase of the project, students will participate in an open school event, where they will present the knowledge, they have gained about how 3D modelling can revolutionize the response to pandemic challenges. This event will include the school community, such as parents, teachers, students, and representatives from relevant companies. This approach creates a valuable opportunity to raise awareness in the local community about the significance of 3D modelling in the healthcare sector.

After the piloting of the educational scenario in the current year within educational hubs, these will be evaluated and refined.  Results will be presented and discussed, and novel iterations will be proposed. Also, we will acknowledge other educational environments as well, such as rural versus urban schools or different economic backgrounds.

**Acknowledgments**

This project has received funding from the European Unions' Horizon 2020 research and innovation programmed under grant agreement No 101006468.

Disclaimer: The sole responsibility for the content on this publication lies with the authors. It does not necessarily reflect the opinion of the European Research Executive Agency (REA) or the European Commission (EC).